\documentstyle[11pt]{article}
\date{}
\parskip=\baselineskip
\begin{document}

\vspace*{1.5 cm}
\begin{center}
{\Large\bf The Sixth Catalogue of Fundamental Stars (FK6)\\[1ex]
and the Problem of Double Stars}\\
\end{center}

\vspace*{0.5 cm}
\begin{center}
{\bf R. Wielen, H. Schwan, C. Dettbarn, H. Jahrei{\ss}, H. Lenhardt}\\[2ex]
Astronomisches Rechen-Institut,\\
M\"onchhofstra{\ss}e 12-14, D\,--\,69120 Heidelberg, Germany\\
\end{center}

\vspace*{1.0cm}
\noindent
{\bf 1. Introduction}

The Sixth Catalogue of Fundamental Stars (FK6) combines the ground-based
astrometric data of the basic fundamental stars, obtained over more than two
centuries and summarized  in the FK5 (Part I: Fricke et al. 1988), with the
observations of the HIPPARCOS astrometry satellite (ESA, 1997). This
combination provides the most accurate proper motions available at present. As
we shall discuss below, the proper motions given in the FK6 are significantly
better than both the proper motions provided by the FK5 or by HIPPARCOS alone.

There are two main reasons why the ground-based observations are able to
improve the HIPPARCOS proper motions considerably: (1) The lower positional
accuracy of the ground-based observations is compensated by a much longer
period of observations, covering more than 200 years for most FK6 stars; (2)
Due to undetected astrometric binaries, the HIPPARCOS proper motions, measured
`instantaneously' during a period of about 3 years only, can deviate
significantly from the long-term proper motions, while the ground-based results
of the FK5 provide already fairly `time-averaged' data. Hence the
ground-based results allow to identify and to correct, at least partially,
`cosmic errors' in the HIPPARCOS proper motions.

The very accurate proper motions of the FK6 can be used either astrophysically,
e.g. for purposes of galactic kinematics, or astrometrically, e.g. for
improving the accuracy of the prediction of stellar positions.

With increasing measuring accuracy, double (and multiple) stars cause
increasingly larger problems in astrometry. This is not only true for objects
whose binary nature is already known but also for undetected binaries. The
uncomfortable consequence for the user is that the  FK6 has to give a variety
of astrometric solutions for the astrometric parameters (proper motion,
position) of a star. The accuracy of and the choice between these solutions
depend on the (often unknown) actual nature of a star (single or double) and on
the model assumed for the motion of the star. The problems of this situation
are discussed by Wielen (1997). The method of `statistical astrometry' is able
to handle quantitatively the astrometric consequences of undetected binaries in
stellar ensembles.

\noindent
{\bf 2. Truly Single Stars}

For a truly single star, we should use the simplest model: the star moves
linearly in time on a straight line in space. The combination of the FK5 data
with the HIPPARCOS observations is then rather straight-forward.

Before combining the two catalogues, we have to reduce the FK5 to the system of
the HIPPARCOS Catalogue. This is done successfully by the analytical method
described by Bien et al. (1977). Consequently, the resulting FK6 is on the
HIPPARCOS system (i.e. on the ICRS).

Let us call the position in one coordinate (e.g. in declination $\delta$) $x$,
the corresponding proper motion $\mu$, the mean epoch (at which $x$ and $\mu$
are uncorrelated) $T$. We use the index $F$ for the FK5 (in the HIPPARCOS
system), and $H$ for HIPPARCOS. From the two positions, $x_F(T_F)$ and
$x_H(T_H)$, we can derive a third proper motion $\mu_0$ (Wielen 1988), in
addition to $\mu_F$ and $\mu_H$:\\
\begin{equation}
\mu_0 = \frac{x_H\,(T_H) - x_F\,(T_F)}{T_H - T_F} \hspace*{0.5cm} .\\[1ex]
\end{equation}
The mean error of $\mu_0$ is given by
\begin{equation}
\varepsilon_{\mu, 0} = \frac{(\varepsilon^2_{x, F, ind} + \varepsilon^2_{x, F,
                      sys} + \varepsilon^2_{x, H})^{1/2}}{T_H - T_F}
\hspace*{0.5cm}
.\\[1ex]
\end{equation}
$\varepsilon_{x, H}$ is the mean measuring error of $x_H(T_H)$,
$\varepsilon_{x, F, ind}$ is the random (`individual') error of
$x_F(T_F)$,
and $\varepsilon_{x, F, sys}$ is the mean error of the systematic difference
in position between the FK5 system and the HIPPARCOS system at the epoch $T_F$
and at the position (and magnitude) of the star under consideration. In other
words, $\varepsilon_{x, F, sys}$ describes the local uncertainty with which the
FK5 system of positions can be reduced to the HIPPARCOS system (It is not the
amount of the systematic difference itself).

If we neglect for a moment the fact that the HIPPARCOS astrometric results for
a star are correlated, we can use a rather direct method for combining the FK5
and the HIPPARCOS data: The resulting FK6 proper motion $\mu_{FK6}$ is the
weighted average of the three proper motions $\mu_0, \mu_F,$ and $\mu_H$. The
weights of the quantities are the inverse squares of their mean errors,
1/$\varepsilon^2_{\mu, 0}$, 1/$\varepsilon^2_{\mu, F}$, 1/$\varepsilon^2_{\mu,
H}$\,. Into $\varepsilon_{\mu, F}$ we have to include the uncertainty of the
reduction of the FK5 proper motion system to the HIPPARCOS system. The weight
of $\mu_{FK6}$ is the sum of the weights of $\mu_0, \mu_F,$ and $\mu_H$\,.
Similarly, the mean epoch $T_{FK6}$ of the star in the FK6 is the weighted
average of $T_F$ and $T_H$, and the mean position $x_{FK6}(T_{FK6})$ is
the weighted average of $x_F(T_F)$ and $x_H(T_H)$, using the weights of
the positions in both cases. Again, the weight of $x_{FK6}$ is the sum of the
weights of $x_F$ and $x_H$. This simple averaging method gives already quite
accurate results which can be understood rather easily. The averaging method
is described in more details in another paper (Wielen et al. 1998). In the
actual FK6, we use a strict method, which takes also care of the correlations
between the HIPPARCOS values of the astrometric parameters (including the
parallax) of a star.

\begin{table}[t]
\vspace*{1.50 cm}
\begin{center}
Table 1\\[1ex]
Error budget for FK6 proper motions in the `single-star mode'\\[-1ex]
\end{center}
\begin{center}
\begin{tabular}{lcc}
\hline\\[-1.0ex]
\multicolumn{3}{c}{Typical mean errors of proper motions}\\
\multicolumn{3}{c}{(in one component, averaged over $\mu_{\alpha\ast}$ and
$\mu_\delta$; units: mas/year)}\\[1.5ex]\hline\\[-2.5ex]
                              & rms average & median\\[0.5ex]\hline\\[-1.8ex]
HIPPARCOS                     & 0.82        & 0.63\\[1.5ex]
FK5\\
\hspace*{0.5cm}random         & 0.76        & 0.64\\
\hspace*{0.5cm}system         & 0.28        & 0.24\\
\hspace*{0.5cm}total          & 0.81        & 0.70\\[1.5ex]
$\mu_0$ (total)               & 0.58        & 0.49\\[1.5ex]
FK6                           & 0.35        & 0.33\\[1.5ex]
ratio of HIPP. to FK6 errors  & 2.3\hphantom{0} &
                                    1.9\hphantom{0}\\[0.5ex]\hline
\end{tabular}
\end{center}
\vspace{-2.00 cm}
\end{table}

We call the resulting solution for the proper motion and position of a star in
the FK6 the `single-star mode', because it is strictly valid for truly single
stars only. In Table 1, we present the typical error budget for the
determination of an FK6 proper motion in this single-star mode. The mean errors
given in Table 1 are root-mean-squared (rms) averages over the individual
values of all the FK6 stars or, for comparison, the median of the individual
values. They are valid for one component, but are averaged over
$\mu_{\alpha\ast}$ and $\mu_\delta$.

From Table 1, we see the following: (1) The proper motion $\mu_0$ is typically
more accurate than $\mu_F$ and $\mu_H$. (2) The final accuracy of the FK6
proper motions in the single-star mode is about two times better than both the
accuracy of the proper motions in the FK5 or in the HIPPARCOS Catalogue. This
significant gain in accuracy justifies the compilation of the FK6 quite well,
even for truly single stars.

\noindent
{\bf 3. Apparently Single Stars and Hidden Binaries}

In real life, we can never be completely confident that a star is truly single.
At best, we can collect a sample of `apparently single stars'. For stars in
such a sample, we have either no indication for a binary nature of the object
at all, or the known binary nature should not affect the astrometric
parameters. Hence we may include e.g. very close spectroscopic binaries, stars
with very distant companions, double stars with known orbits for which the
center-of-mass motion is given etc.

In other papers (Wielen 1995, Wielen et al. 1997), we have shown that the
HIPPARCOS proper motions $\mu_H$ in such a sample of `apparantly single stars'
suffer from cosmic errors, due to undetected astrometric binaries. The
basic fundamental stars show the effect of cosmic errors most clearly, because
of the small measuring errors in their proper motions and positions. The mean
cosmic error
$c_\mu$ in $\mu_H$, corresponding to ($\eta$(0))$^{1/2}$ in the
terminology of Wielen (1997), was determined from a comparison of $\mu_F$ and
$\mu_H$ in a sample of 1202 apparently single FK stars as
$c_\mu$ = 2.13 mas/year.

\begin{table}[t]
\vspace*{1.50 cm}
\hspace*{-1.00 cm}
\begin{center}
Table 2\\[1ex]
Cosmic error $c_\mu$ in HIPPARCOS proper motions\\
\end{center}
\begin{center}
\vspace*{-0.30 cm}
\hspace*{-1.00 cm}
\begin{tabular}{lccccccc}\hline\\[-1.0ex]
Quantity \hspace*{1.48cm}{Catalogue:} & FK5      & FK5   & FK4     & FK3   &
                                                                    GC & GC\\
\hspace*{\fill}{Sample of stars:}
& 1202 FK  & 1202 FK & 1202 FK & 1202 FK & 1201 FK &
                                                                 11\,773 GC\\
\hspace*{\fill}{rms of:} & $\mu_F-\mu_H$ & $\mu_0-\mu_H$ & $\mu_0-\mu_H$ &
$\mu_0-\mu_H$ & $\mu_0-\mu_H$ & $\mu_0-\mu_H$\\[1.5ex]\hline\\[-1.5ex]
rms difference $\mu_F$ or $\mu_0 - \mu_H$  & 2.38 & 2.04 & 2.07 & 2.12 & 2.18 &
                                              2.63\\[1ex]
measuring errors:\\
\hspace*{0.5cm} $\varepsilon_{\mu, H}$ & 0.68 & 0.68 & 0.68 & 0.68 & 0.68 &
                                                                        0.76\\
\hspace*{0.5cm} $\varepsilon_{\mu, F}$ or $\varepsilon_{\mu, 0}$
\hspace*{1.00 cm} (total)                  & 0.82 & 0.59
                                            & 0.55 & 0.61 & 0.67 & 1.43\\[1ex]
remaining part =\\
cosmic error $c_\mu$ & 2.13 & 1.83 & 1.88 & 1.91 & 1.96 & 2.07\\[1ex]
mean epoch difference\\
$T_{cat} - T_H$ [years] & 42\hphantom{.0} & 42\hphantom{.0} & 75
\hphantom{0} & 88\hphantom{.0} & 89 \hphantom{0} & 91\hphantom{0}\\[1ex]
median of $m_V$ [mag] & 4.9\hphantom{0} & 4.9\hphantom{0} & 4.9\hphantom{0} &
    4.9\hphantom{0} & 4.9\hphantom{0} & 6.8\hphantom{.}\\[1ex]
median of parallax [mas] & 11.0\hphantom{0,} & 11.0\hphantom{0,} &
11.0\hphantom{0,} & 11.0\hphantom{0,} & 11.0\hphantom{0,} & 6.8\hphantom{.}\\
[0.8ex]\hline\\
\end{tabular}
\end{center}
\vspace*{-0.50 cm}
\hspace*{-0.70 cm}
Note: The units of $\mu, \, \varepsilon$, and $c_\mu$ are mas/year.\\
\vspace{-2.00 cm}
\end{table}

%
%
\pagebreak

To confirm this value of $c_\mu$,
we have compared $\mu_H$ also with $\mu_0$ for
the same sample of FK stars. The proper motion $\mu_0$ has been obtained from
combinations of the HIPPARCOS observations with the following compilation
catalogues: FK5 (Fricke et al. 1988), FK4 (Fricke et al. 1963), FK3 (Kopff
1937, 1938), and General Catalogue GC (Boss et al. 1937).
The cosmic error $c_\mu$ is derived by subtracting the measuring errors
$\varepsilon_{\mu, H}$
and $\varepsilon_{\mu, F}$(total)
or $\varepsilon_{\mu, 0}$(total)
of $\mu_H$ and of $\mu_F$ or $\mu_0$ quadratically from the rms difference
$\mu_F-\mu_H$ or $\mu_0-\mu_H$.
Table 2 shows that
the resulting values of the cosmic error
$c_\mu$ are in good agreement, both among
the various catalogues and with the value of
$c_\mu$ quoted above and derived from
a comparison of $\mu_{FK5}$
with $\mu_H$. The advantage of using $\mu_0$, instead
of the proper motion given in the catalogue itself, is the good accuracy of
$\mu_0$ even for old compilation catalogues. The use of e.g. $\mu_{GC}$, with
typical measuring errors of the order of 10 mas/year, would not allow a
meaningful determination of $c_\mu$\,$\sim$\,2 mas/year. It is especially
remarkable
that the much larger sample of about 12\,000 well-measured and apparently
single GC stars gives essentially the same typical value for c as we have
determined earlier from the FK stars.

We have finally adopted for the FK6 the overall value of 2.13 mas/year for
$c_\mu$.
In detail, we are using in the FK6 a cosmic error
$c_\mu(r)$ which depends on the
distance $r$ of the star from the Sun, determined empirically from the run of
$c_\mu$ with $r$ among the 1202 FK stars.
The distance $r$ is obtained usually from
the HIPPARCOS parallax. For small and uncertain parallaxes, we use photometric
distances if these are expected to be more accurate. We adopt also a cosmic
error of 10 mas in the HIPPARCOS positions $x_H(T_H)$.

In the presence of cosmic errors  in the HIPPPARCOS data, we determine, in
contrast to the single-star mode of the FK6, now a `long-term prediction' in
the FK6. For the long-term prediction, we modify the method of combining the
FK5 data with the HIPPARCOS observations, described in Section 2, in the
following way: (a) A total mean error $\varepsilon_{\mu, H}$ of $\mu_H$ is
calculated by adding quadratically the measuring error of $\mu_H$ and the
cosmic error
$c_\mu$ of $\mu_H$; (b) Similarly, the mean error of $\mu_0$ is
increased by taking the cosmic error of $x_H(T_H)$ into account; (c) The
cosmic error in $x_H(T_H)$ is also added to the measuring error of a
HIPPARCOS position. After all these modifications of some of the mean errors
and hence some of the weights, the methods of Section 2 are applied. This
procedure is mathematically justified by the methods derived within the scheme
of statistical astrometry (Wielen 1997) for the case that $\mu_F$ is a mean
(long-term averaged) proper motion without any cosmic error. While this
assumption is certainly not strictly true, it should produce reasonable
results.

\begin{table}[t]
\vspace*{1.50 cm}
\begin{center}
Table 3\\[1ex]
Error budget for FK6 proper motions in the `long-term prediction mode'\\[-1ex]
\end{center}
\begin{center}
\begin{tabular}{lcc}\hline\\[-1.0ex]
\multicolumn{3}{c}{Typical mean errors of proper motions}\\
\multicolumn{3}{c}{(in one component, averaged over $\mu_{\alpha\ast}$ and
                     $\mu_\delta$; units: mas/year)}\\[1.5ex]\hline\\[-2.5ex]
                              & rms average & median\\[0.5ex]\hline\\[-1.8ex]
HIPPARCOS\\
\hspace*{0.5cm}measuring error & 0.68       & 0.60\\
\hspace*{0.5cm}cosmic error    & 2.13       & (2.04)\\
\hspace*{0.5cm}total           & 2.24       & (2.13)\\[1.5ex]
FK5\\
\hspace*{0.5cm}random          & 0.77       & 0.66\\
\hspace*{0.5cm}system          & 0.28       & 0.24\\
\hspace*{0.5cm}total           & 0.82       & 0.70\\[1.5ex]
$\mu_0$ (total)                & 0.59       & 0.51\\[1.5ex]
FK6                            & 0.49       & 0.45\\[1.5ex]
ratio of HIPP. (total) to FK6 errors & 4.6\hphantom{0}  &
                                             (4.7)\hphantom{.}\\[0.5ex]\hline
\end{tabular}
\end{center}
\vspace{-2.00 cm}
\end{table}

In Table 3, we give the error budget of the FK6 proper motions in the long-term
prediction mode. If we compare the accuracy of these FK6 long-term proper
motions with the accuracy of HIPPARCOS proper motions, now taking the cosmic
error
$c_\mu$ in $\mu_H$ into account, we see that the FK6 is more accurate than
HIPPARCOS by a factor of more than 4. This is certainly a remarkable
improvement.

The long-term proper motions of the FK6 should be preferred over the
instantaneous HIPPARCOS proper motions both for galactic kinematics as well as
for long-term predictions of stellar positions. As soon as the epoch of the
predicted position of a star differs by more than about 10 years from $T_H$ =
1991.25, the long-term prediction of the FK6 is more accurate and gives more
reliable error estimates than the direct use of the HIPPARCOS Catalogue.

For epochs close to $T_H$, one may use either directly the instantaneous
HIPPARCOS data or the `short-term prediction mode' of the FK6. This short-term
prediction is derived by a modification of the methods described in Section 2
in a way formally similar to that used for the long-term prediction. The only
difference is that in the short-term prediction, the cosmic errors have to be
added to the FK5 mean errors of $\mu_F$ and $x_F(T_F)$, instead of the
addition to the HIPPARCOS ones.

\noindent
{\bf 4. Known Double Stars}

The basic FK5 contains 1535 stars. For 302 of these stars (20 percent), special
solutions (C, O, G, X, V) are given in the HIPPARCOS Catalogue, indicating
their binary nature. Other binaries known from ground-based observations,
especially visual binaries with magnitude differences of more than about 3, are
treated by HIPPARCOS as single stars (standard solutions).

For most of the known double stars, the combination of FK5 and HIPPARCOS data
is usually not straightforward. Many different situations require rather
different operations. In most cases, individual `corrections' have to be
applied to the data, either to FK5 or to HIPPARCOS, before we can combine the
data into the FK6. Even then, the results for these binaries are often rather
uncertain and of doubtful significance. In this paper, we can indicate a few of
these problems only.

The first problem to be solved for binaries is usually to find a common
astrometric `reference point' for the FK5 and HIPPARCOS which is also
meaningful for the user of the FK6. Good reference points are, for example, the
center-of-mass of a binary, or the positions of the components A and B of a
double star with purely linear relative motions. These reference points can be
combined like single stars.

Other reference points are less well-defined. This is especially true for the
photo-center of A and B. The photo-center depends on the photometric system (V
or Hp), and implicitly on the ratio between the orbital period and
the time coverage of observations. For binaries with periods of the order of
a few decades, the FK5 gives essentially the {\it time-averaged} position and
motion of the photo-center, while HIPPARCOS provides the {\it instantaneous}
photo-center. If we know the orbit and the individual magnitudes and colours of
the binary components accurately, we can harmonize the data. But in most cases,
we can do this in a statistical manner only.

For determining the most appropriate reference point for a given double star,
it is very important to know its orbital period $P$. For 47 visual binaries
in the basic FK5, we know $P$ rather accurately from a well-determined relative
orbit. For the remaining majority of cases, we have obtained a statistical
estimate of $P$ from the observed separation, parallax and photometry, using
the mass-luminosity relation. In many cases, these estimated orbital periods
indicate rather clearly whether or not the FK5 or HIPPARCOS provide
`time-averaged' or `instantaneous' data.

In favourable cases, we can obtain very accurate `corrections'. For example,
for many short-period astrometric binaries with small separations, the O
solutions of HIPPARCOS provide the center-of-mass as the reference point. In
these cases, the FK5 data are to be interpreted as valid for the time-averaged
photo-center. Using the known orbital elements, especially the eccentricity, we
can derive the constant shift between the mean photo-center and the
center-of-mass. This shift can then be used to reduce the FK5 position to the
center-of-mass.

For 95 stars (6 per cent) of the basic FK stars, the HIPPARCOS Catalogue
provides non-linear solutions of type G. It can be shown that most of these
stars are astrometric binaries with orbital periods of a few years. For the
stars with G solutions, we neglect the HIPPARCOS proper motions completely,
because they differ from the FK5 proper motions typically by about 10 mas/year.
Hence we combine only the HIPPARCOS positions with the FK5 positions and proper
motions. The resulting FK6 data for the stars with G solutions by HIPPARCOS
refer then to the time-averaged photo-center of these binaries.

\noindent
{\bf 5. Astrometrically Excellent Stars}

Being confronted with the often very nasty problems caused by double stars, one
may be inclined to eliminate at least known double stars entirely from the FK6.
This could be indeed a solution from the purely astrometric point-of-view. In
high-precision astrometry, the accuracy of the astrometric data for double
stars is often by an order of magnitude or more lower than for single stars.
However, for many astrophysical problems one needs at least the proper motion
of a double star, since many interesting objects (e.g. $\delta$ Cep) are
binaries. In that sense, the FK6 aims at deriving the `best' proper motions
also for such known double stars, even if the accuracy is not as good as for
single stars.

As a compromise between the conflicting demands of highest astrometric accuracy
and the inclusion of double stars for completeness, we shall identify in the
FK6 a subsample of `astrometrically excellent stars' by special flags. Such an
astrometrically excellent star should behave essentially like a well-measured
single star.

The selection criteria for astrometrically excellent stars are: (1) No
disturbing duplicity is known, neither from ground-based observations nor from
HIPPARCOS measurements. (2) Good agreement between the three proper motions
$\mu_H, \, \mu_F,$ and $\mu_0$. This agreement indicates, at least
statistically, that the cosmic errors in the HIPPARCOS proper motions for these
stars are small. (3) Small measuring errors of the proper motions.

The number of astrometrically excellent stars depends strongly on how stringent
the selection criteria are chosen. We think that many hundreds of FK6 stars are
qualified as astrometrically excellent stars. Perhaps we will indicate
different levels of excellence by different types of flags, similar to the
numbers of `Michelin stars' for indicating the excellence of a restaurant. Of
course, further observations may show that the star is actually not qualified
as being astrometrically excellent. In any case, the `excellence' attributed to
a star in the FK6 is significant only on the level of accuracy reached at
present. Future space missions may be much more disturbed by the binary nature
of the majority of stars.

\noindent
{\bf 6. Conclusions}

The proper motions of the FK6, and hence the positions predicted from the FK6
data, are expected to be significantly more accurate than the data given in the
HIPPARCOS Catalogue and in the FK5. For truly single stars, the accuracy is
improved by a factor of about 2 by the FK6. For the ensemble of apparently
single stars, in which the HIPPARCOS proper motions suffer from cosmic errors
due to undetected astrometric binaries, the gain is even higher, by a factor
of more than 4.
Double stars pose special problems which are handled rather
individually in the FK6. Among all the FK6 stars, a subset of `astrometrically
excellent stars' is identified for high-precision astrometry, avoiding as far
as possible a disturbing binary nature of the objects.\\

\noindent
{\bf References}
\vspace*{-0.3cm}
\begin{description}
\item Bien, R., Fricke, W., Lederle, T., Schwan, H., 1977, Ver"ff. Astron.
Rechen-Inst. Heidelberg No. 27

\item Boss, B., Albrecht, S., Jenkins, H., Raymond, H., Roy, A.J., Varnum,
W.B., Wilson, R.E., 1937, General Catalogue of 33\,342 Stars for the Epoch
1950, Carnegie Institution of Washington, Publ. No. 486

\item ESA, 1997, The Hipparcos Catalogue, ESA SP-1200

\item Fricke, W., Kopff, A., Gliese, W., Gondolatsch, F., Lederle, T., Nowacki,
H., Strobel, W., Stumpff, P., 1963, Ver"ff. Astron. Rechen-Inst. Heidelberg No.
10

\item Fricke, W., Schwan, H., Lederle, T., Bastian, U., Bien, R., Burkhardt,
G., du Mont, B., Hering, R., J"hrling, R., Jahreiá, H., R"ser, S.,
Schwerdtfeger, H.M., Walter, H.G., 1988, Ver"ff. Astron. Rechen-Inst.
Heidelberg No. 32

\item Kopff, A., 1937, Ver"ff. Astron. Rechen-Inst. Berlin No. 54

\item Kopff, A., 1938, Abh. Preuá. Akad. Wiss., Jahrgang 1938, Phys.-math.
Klasse, No. 3

\item Wielen, R., 1988, in: IAU Symposium No. 133, Mapping the Sky, eds. S.
Debarbat, J.A. Eddy, H.K. Eichhorn, A.R. Upgren, Kluwer Publ. Comp., Dordrecht,
p. 239

\item Wielen, R., 1995, in: Future Possibilities for Astrometry in Space, eds.
M.A.C. Perryman, F. van Leeuwen, ESA SP-379, p. 65

\item Wielen, R., 1997, Astron. Astrophys. 325, 367

\item Wielen, R., Schwan, H., Dettbarn, C., Jahreiá, H., Lenhardt, H., 1997,
in: Hipparcos Venice '97, Presentation of the Hipparcos and Tycho Catalogues
and first astrophysical results of the Hipparcos space astrometry mission, eds.
Battrick, B., Perryman, M.A.C., Bernacca, P.L., ESA SP-402, p. 727

\item Wielen, R., Schwan, H., Dettbarn, C., Jahreiá, H., Lenhardt, H., 1998,
in: Proceedings of the International Colloquium on `Modern Astrometry and
Astrodynamics', held in honor of Heinrich Eichhorn at Wien, Austria, 25-26 May
1998, ed. R. Dvorak, to be published by the ™sterreichische Akademie der
Wissenschaften, Wien
\end{description}

\newpage

\end{document}